\documentclass[12pt,showpacs]{revtex4}
\usepackage[english]{babel}
\usepackage{graphicx}
\usepackage{epsfig}

\def\oggi{\number\day.\space 
\ifcase\month\or
 1.\or 2.\or 3.\or 4.\or 5.\or 6.\or
 7.\or 8.\or 9.\or 10.\or 11.\or 12.\fi
 \space\number\year}

\def\be{\begin{equation}}
\def\ee{\end{equation}}
\def\bea{\begin{eqnarray}}
\def\eea{\end{eqnarray}}

\def\oneh{{\textstyle {1\over 2}}}

\def          
\simeq{{\ \lower2pt\hbox{$-$}\mkern-13mu \raise2pt \hbox{$\sim$}\ }} 

\def           
\slasha#1{{\hbox{$\not$} \mkern-3mu\hbox{$\,#1\,$}}} 

\newcommand{\tvec}[1]{\mbox{\boldmath{$#1$}}}
\newcommand{\svec}[1]{\mbox{\boldmath{$\scriptstyle #1$}}}

\def\dsp{\displaystyle}

\begin{document}


\title{
Nucleon spin densities in a light-front constituent quark model}
\normalsize
\author
{B.~Pasquini and S.~Boffi}
\affiliation{
Dipartimento di Fisica Nucleare e Teorica, Universit\`{a} di Pavia, and\\
Istituto Nazionale di Fisica Nucleare, Sezione di Pavia, I-27100 Pavia, Italy}

\begin{abstract}
{The first moment of longitudinal and transverse spin densities of quarks in the nucleon are calculated in a light-front constituent quark model for the different cases of quark and nucleon polarization. Significant distortions are found for the transverse spin densities. In particular the Sivers function is predicted with opposite sign for up and down quarks and the Boer-Mulders function is predicted large and negative for both up and down quarks, in agreement with lattice calculations. Quite a different spin distribution is obtained for up and down quarks in the cases of quarks and proton transversely or longitudinally polarized in the same direction.
}
\end{abstract}
\date{\today}
\maketitle

At leading order in the deep inelastic regime the quark structure of the nucleon is fully parameterized in terms of generalized parton distributions (GPDs) (for recent reviews, see Refs.~\cite{Diehlrep,Jiarnps,BR05}). Four of them, usually indicated by $H$, $E$, $\tilde H$, $\tilde E$, are helicity conserving (chiral even) and the other four, denoted by $H_T$, $\tilde H_T$, $E_T$, $\tilde E_T$, are helicity flip (chiral odd) distributions~\cite{Hoodbhoy98a,Diehl01}. They are all functions of three kinematical variables, i.e. the longitudinal momentum fraction $x$ of the quark, the invariant momentum transfer $t$ and the skewness parameter $\xi$ describing the fraction of the longitudinal momentum transfer to the nucleon. For each flavour they are related to important quantities characterizing the nucleon structure. In the forward limit, i.e. $t=0$, $H$, $\tilde H$ and $H_T$ reduce to the parton density $f_1(x)$, helicity distribution $g_1(x)$ and transversity distribution $h_1(x)$, respectively. First moments of $H$, $E$, $\tilde H$, and $\tilde E$ are identified with the quark contribution to the Dirac and Pauli form factors, $F_1$ and $F_2$, and the axial and pseudoscalar form factors, $G_A$ and $G_P$, respectively. Thus in the forward limit  $E$ and $\tilde H$ are related to the quark contribution to the anomalous magnetic moment $\kappa$ and the axial-vector  coupling $g_A$. The first moment of the combination $E_T + 2\tilde H_T$ in the forward limit gives the quark contribution to the anomalous tensor magnetic moment $\kappa_T$. Finally, $\tilde E_T$ vanishes for $\xi=0$. 

When $\xi=0$ and $x>0$, by a two-dimensional Fourier transform to impact parameter space GPDs can be interpreted as densities of quarks with longitudinal momentum fraction $x$ and transverse location $\tvec b$ with respect to the nucleon center of momentum~\cite{Burkardt00a,Burkardt03}. Depending on the polarization of both the active quark and the parent nucleon, according to 
Refs.~\cite{Burkardt03,diehlhagler05} one defines three-dimensional densities $\rho(x,{\tvec b}, \lambda,\Lambda)$ and  $\rho(x,{\tvec b},{\tvec s},{\tvec S})$ representing the probability  to find a quark with longitudinal momentum fraction $x$ and transverse position $\tvec b$ either with light-cone helicity $\lambda$ ($=\pm 1$) in the nucleon with longitudinal polarization $\Lambda$ ($=\pm 1$) or with transverse spin $\tvec s$ in the nucleon with transverse spin $\tvec S$. They read
\be
\rho(x,{\tvec b}, \lambda,\Lambda) =  \oneh \left[ H(x,{b}^2) 
  + b^j\varepsilon^{ji} S^i  \frac{1}{M}\, 
       E'(x,{b}^2)
  + \lambda \Lambda \tilde{H}(x,{b}^2) \,\right] ,
 \label{eq:long}
 \ee
\bea
\rho(x,{\tvec b},{\tvec s},{\tvec S}) 
&= &{}\dsp \oneh\left[ H(x,{b}^2)  + s^iS^i\left( H_T(x,{b}^2)  -\frac{1}{4M^2} \Delta_b \tilde H_T(x,{b}^2) \right) \right.
\nonumber\\
& &\quad{}\dsp + \frac{b^j\varepsilon^{ji}}{M}\left(
S^iE'(x,{b}^2)  + s^i\left[ E'_T(x,{b}^2)  + 2 \tilde H'_T(x,{b}^2) \right]\right)
\nonumber\\
& &\quad\left.{}\dsp+ s^i(2b^ib^j - b^2\delta_{ij}) S^j\frac{1}{M^2} \tilde H''_T(x,{b}^2) \right] .
 \label{eq:transv}
 \eea
The distributions $H$, $E$, $\tilde H$, $H_T$, etc. are the Fourier transform of the corresponding GPDs, i.e.
\be
f(x,{b}^2) = \int\frac{d^2{\tvec\Delta}}{(2\pi)^2} \,e^{-i{\svec b}\cdot{\svec\Delta}} \,f(x,\xi=0,t=-{\tvec\Delta}^2),
\label{eq:fourier}
\ee
where $\tvec\Delta$ is the transverse momentum transfer to the nucleon. As such, they depend on ${\tvec b}$ only via its square ${\tvec b}^2=b^2$ thanks to rotation invariance. In Eqs.~(\ref{eq:long}) and (\ref{eq:transv}) the shorthand notations
\be
f' = \frac{\partial}{\partial b^2}\, f ,
\qquad
f''= \Big( \frac{\partial}{\partial b^2} \Big)^2 f,
\qquad
\Delta_b f
= \frac{\partial}{\partial b^i}\, \frac{\partial}{\partial b^i}\, f
= 4\, \frac{\partial}{\partial b^2}
    \Big( b^2 \frac{\partial}{\partial b^2} \Big) f 
\label{eq:deriv}
\ee
have been used, and the two-dimensional antisymmetric tensor $\varepsilon^{ij}$ has been introduced with $\varepsilon^{12} = -\varepsilon^{21} = 1$ and $\varepsilon^{11} = \varepsilon^{22} = 0$. $M$ is the nucleon mass, and Roman indices are to be summed over.

In Eq.~(\ref{eq:long}) the first term with $H$ describes the density of unpolarized quarks in the unpolarized proton. The term with $E'$ introduces a sideways shift in such a density when the proton is transversely polarized, and the term with $\tilde H$ reflects the difference in the density of quarks with helicity equal or opposite to the proton helicity. 

In the three lines of Eq.~(\ref{eq:transv}) one may distinguish the three contributions corresponding to monopole, dipole and quadrupole structures. The unpolarized quark density $\oneh H$ in the monopole structure is modified by the chiral-odd terms with $H_T$ and $\Delta_b \tilde H_T$ when both the quark and the proton are transversely polarized. Responsible for the dipole structure is either the same chiral-even contribution with $E'$ from the transversely polarized proton appearing in the longitudinal spin distribution~(\ref{eq:long}) or the chiral-odd contribution with $E'_T+2\tilde H'_T$ from the transversely polarized quarks or both. The quadrupole term with $\tilde H''_T$ is present only when both quark and proton are transversely polarized.

Lattice calculations accessing the lowest two $x$-moments of the transverse 
spin densities of quarks in the nucleon have recently been
 presented~\cite{QCDSF06a}, and
 impact parameter dependent parton distributions in phenomenological models of
 hadron 
light-cone wave functions (LCWFs) have been studied in Ref.~\cite{DMR07}.

\begin{figure}[h]
\begin{center}
\includegraphics[width=16 truecm]{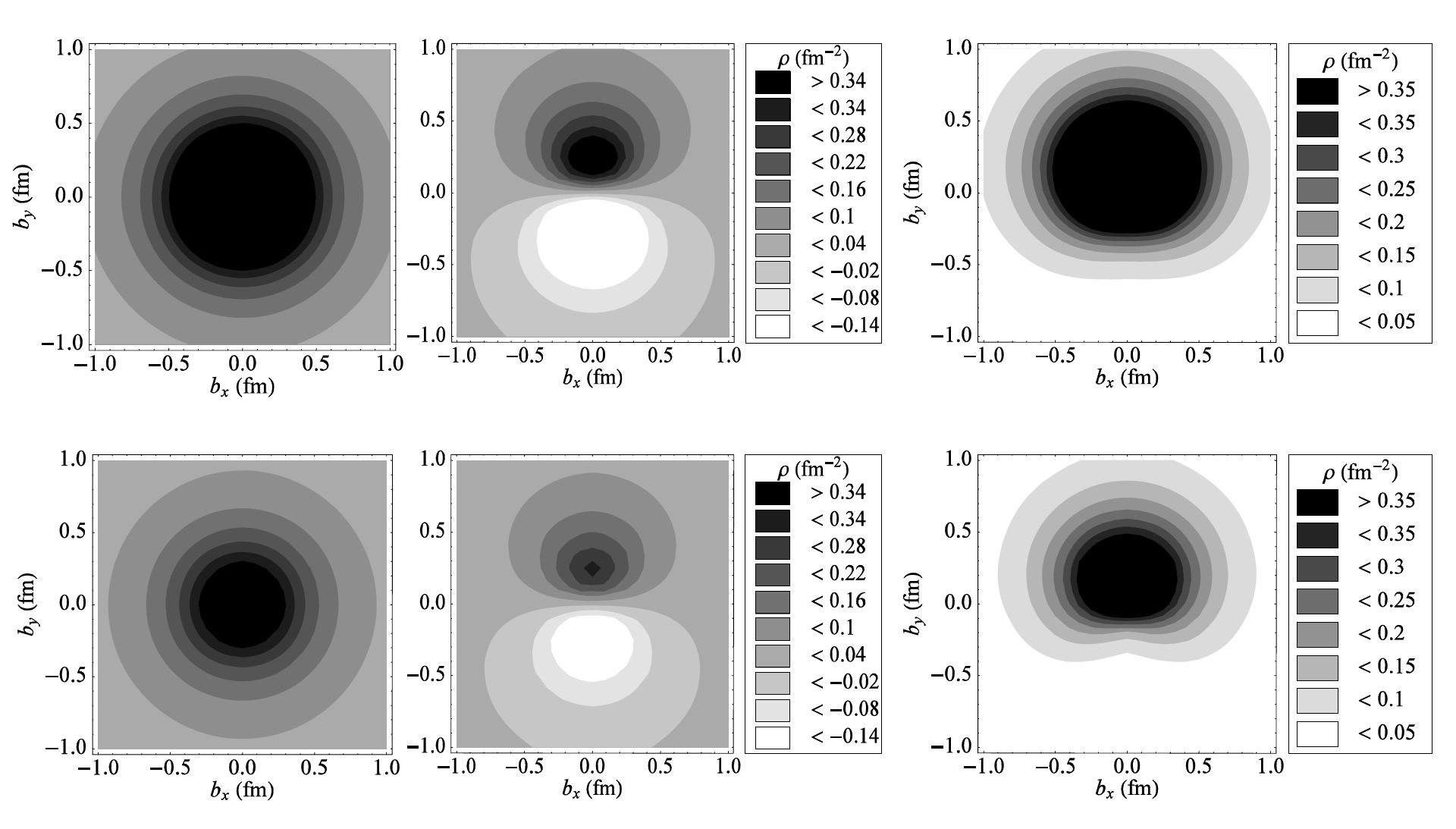}
\end{center}
\caption{\small The monopole contribution $\frac{1}{2}H$ (left) for unpolarized quarks, the dipole contribution $-\frac{1}{2} s_x b_y (E'_T+2\tilde H'_T)/M$ (middle) for (transversely) $\hat x$-polarized quarks, and the sum of both (right) in an unpolarized proton. The upper (lower) row gives the results for up (down) quarks.}
\label{fig:fig1}
\end{figure}

In this Letter the first $x$-moments of the spin distributions
\be
\rho({\tvec b}, \lambda,\Lambda) =\int dx\, \rho(x,{\tvec b}, \lambda,\Lambda), \quad
 \rho({\tvec b},{\tvec s},{\tvec S}) = \int dx\,  \rho(x,{\tvec b},{\tvec s},{\tvec S})
 \label{eq:moments}
\ee
are studied as functions of the transverse position and different quark and 
proton polarizations taking advantage of the overlap representation of LCWFs
 that was originally proposed 
in Refs.~\cite{DFKJ01,BDH01} and successfully applied 
to GPDs~\cite{BPT03,BPT04,PPB05,PPB06a,PPB06b}. 
\begin{figure}[t]
\begin{center}
\includegraphics[width=16 cm]{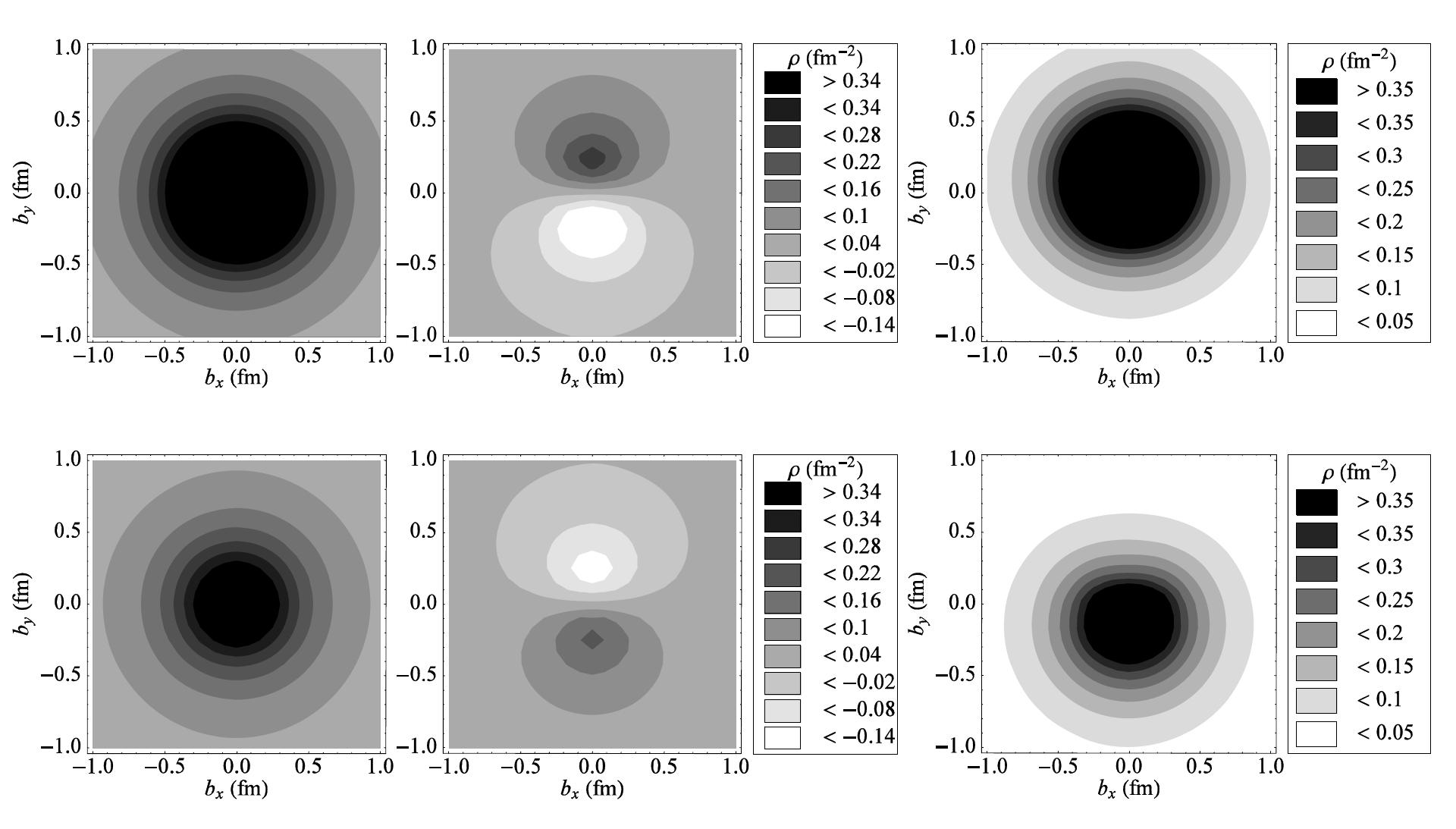}
\end{center}
\caption{\small With unpolarized quarks the monopole contribution $\frac{1}{2}H$ (left) for an unpolarized proton, the dipole contribution $-\frac{1}{2} S_x b_y E'/M$ for a (transversely) $\hat x$-polarized proton, and the sum of both (right). The upper (lower) row gives the results for up (down) quarks.}
\label{fig:fig2}
\end{figure}
In such an approach the lowest-order Fock-space components of LCWFs with three valence quarks are directly linked to wave functions derived in constituent quark models. Assuming SU(6) symmetry and separating  the spin-isospin component $\Phi(\{\lambda_i\},\{\tau_i\})$ from the momentum-space part $\psi(\{{\tvec k}_i\})$ of the wave function, the eigenfunctions  $\Psi(r,\{\lambda_{i}\},\{\tau_i\})$ of the light-front Hamiltonian of the nucleon described as a system of three interacting quarks (with helicity $\lambda_i$ and isospin $\tau_i$, $i=1,2,3$) are written as
\bea
\label{eq:transform}
& & \Psi (r,\{\lambda_{i}\},\{\tau_i\})
=  2(2\pi)^3\left[\frac{1}{M_0}\frac{\omega_1\omega_2\omega_3} {y_1y_2y_3}\right]^{1/2} 
\psi(\{{\tvec k}_i\})
\nonumber\\
& & \qquad{} \times\sum_{\{\lambda'_i\}}
{D}^{1/2\,*}_{\lambda'_{1}\lambda_{1}}(R_{cf}(\tvec k_1))
{D}^{1/2\,*}_{\lambda'_{2}\lambda_{2}}(R_{cf}(\tvec k_2))
{D}^{1/2\,*}_{\lambda'_{3}\lambda_{3}}(R_{cf}(\tvec k_3))
\,
\Phi_{\lambda,\tau}(\{\lambda'_i\},\{\tau_i\}) ,
\\ \nonumber
\eea
\noindent
where $r$ collectively denotes the set of light-cone momentum variables $\{\tvec k_i=(y_i,{\tvec k_\perp}_i)\}$ of the involved quarks, $M_0$ is the mass of the non-interacting three-quark system, $\omega_i=(k^+_i+k^-_i)/\sqrt{2}$, and the matrix ${D}^{1/2}_{\lambda\mu}(R_{cf}(\tvec k))$  is given by the representation of the Melosh rotation $R_{cf}$ in spin space transforming the canonical spin into the light-front spin. Here the momentum space wave function $\psi(\{{\tvec k}_i\})$ has the same power-law behaviour with the same parameters used in the relativistic quark model of Ref.~\cite{Schlumpf94a}. As such the original (instant form) wave function only contains $S$-wave components.

The various GDPs are then calculated with the above LCWFs as explained in Refs.~\cite{BPT03,BPT04,PPB05}, Fourier transformed according to Eq.~(\ref{eq:fourier}) and integrated over $x$ to obtain the first moments (\ref{eq:moments}).

With such a model only valence quarks are considered. Therefore, the integrals in Eq.~(\ref{eq:moments}) are restricted to $0\le x\le 1$. In order to have a first idea of how spin distributions look like in a nucleon this limitation is not dramatic because valence quarks are known to dominate at large and intermediate $x$ ($x\ge 0.2$).

\begin{figure}[h]
\begin{center}
\includegraphics[width=16 cm]{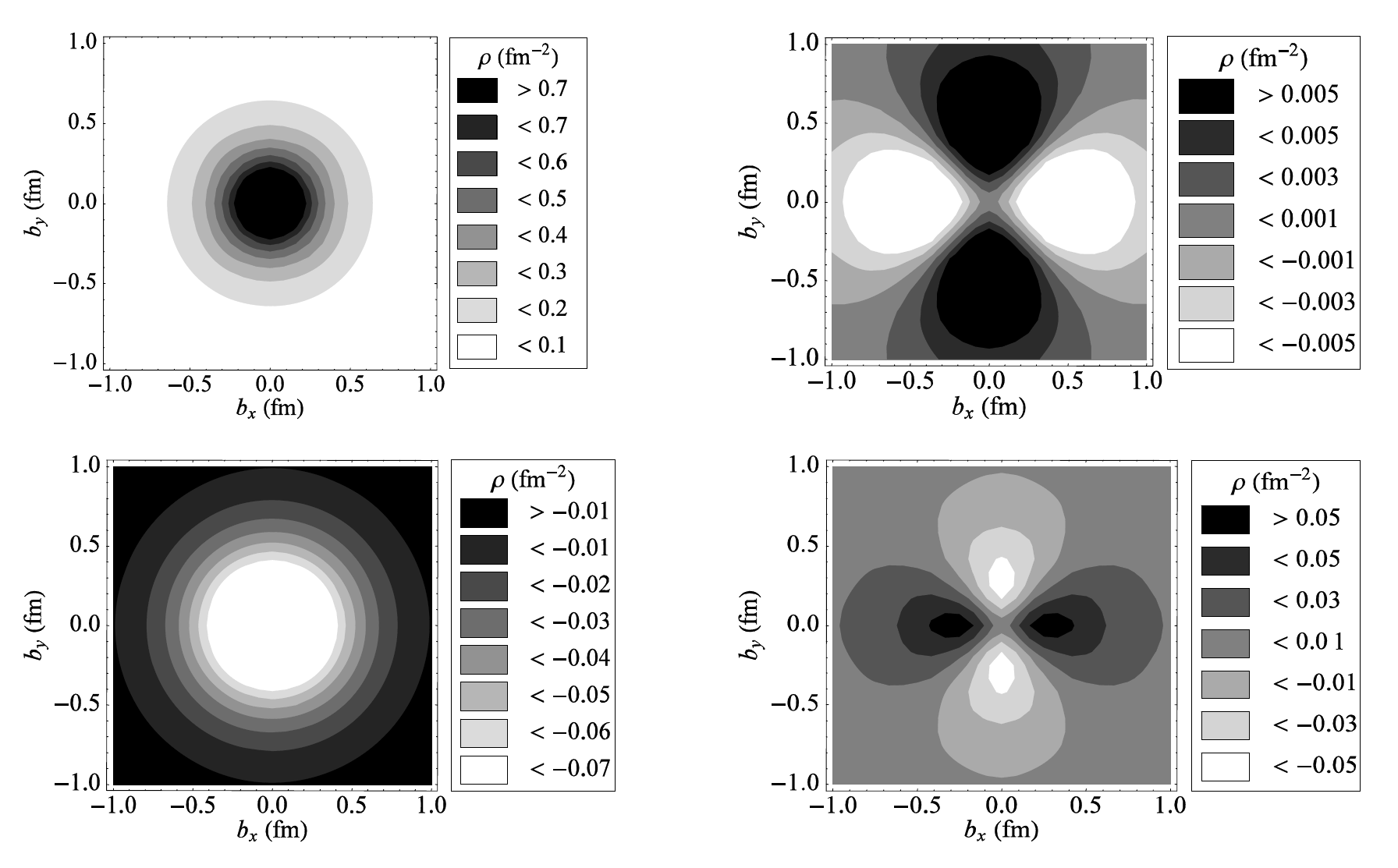}
\end{center}
\caption{\small The monopole contribution $\oneh s_xS_x(H_T-\Delta_b\tilde H_T/4M^2)$ (left) and the quadrupole contribution $\oneh s_xS_x(b_x^2-b_y^2)\tilde H''_T/M^2$ (right) for $\hat x$-polarized quarks in a nucleon also polarized along $\hat x$. The upper (lower) row gives the results for up (down) quarks.}
\label{fig:fig3}
\end{figure}

\begin{figure}[h]
\begin{center}
\includegraphics[width=16 cm]{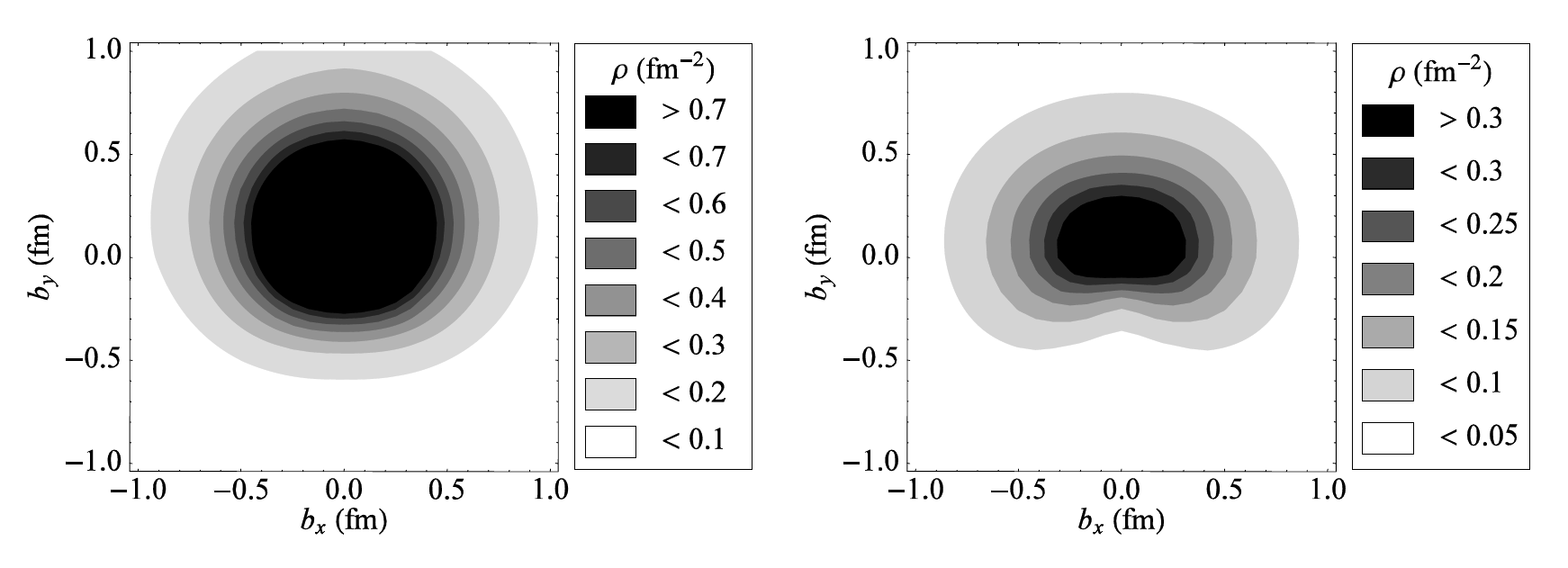}
\end{center}
\caption{\small The total spin distribution as a sum of monopole, dipole and quadrupole terms,  for $\hat x$-polarized quarks in a proton also polarized along $\hat x$; left (right) panel for up (down) quarks.}
\label{fig:fig4}
\end{figure}

\begin{figure}[h]
\begin{center}
\includegraphics[width=16 cm]{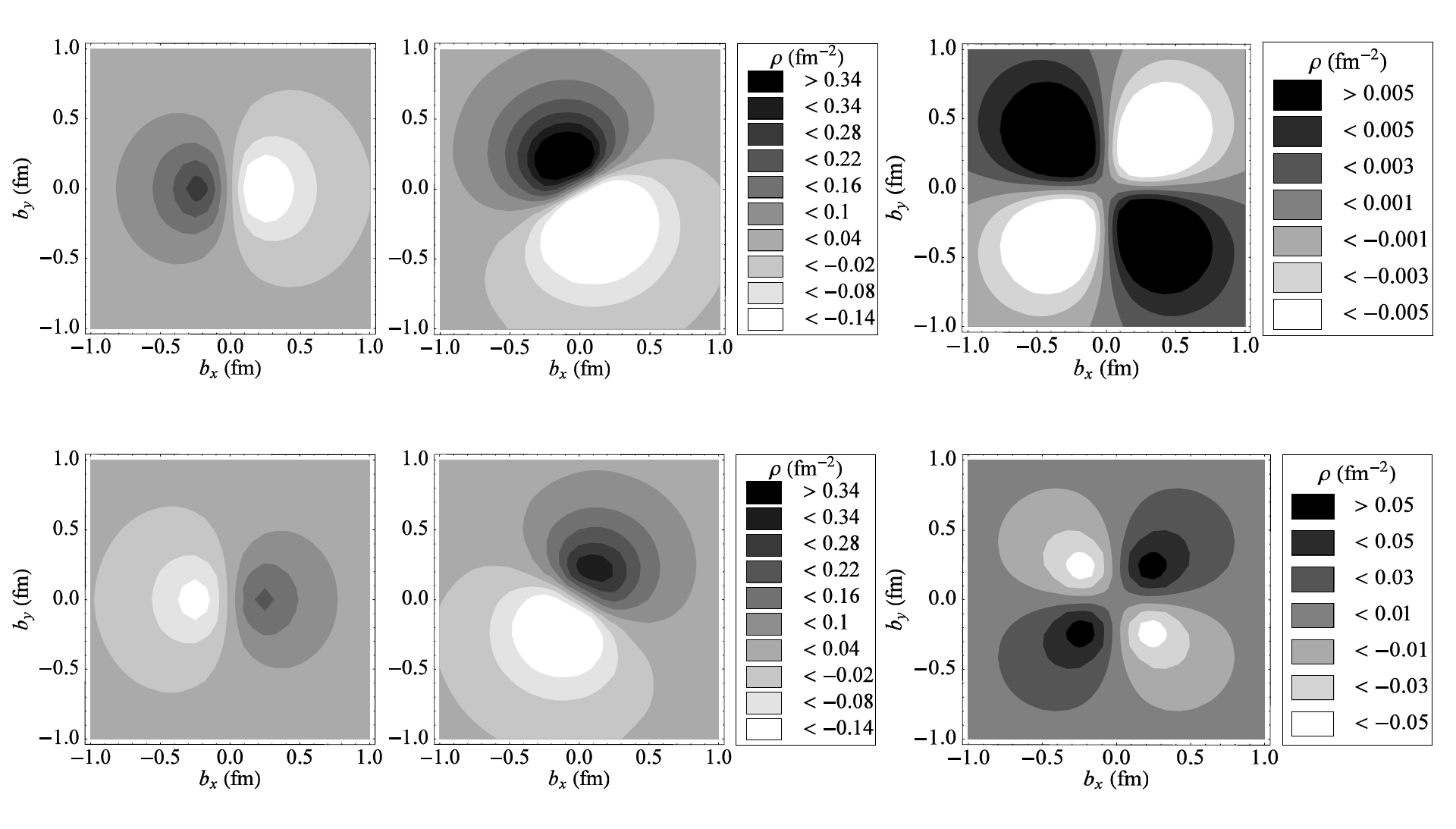}
\end{center}
\caption{\small  The dipole contribution $\oneh S_yb_xE'$ (left), the total dipole contribution $\oneh[S_yb_xE'-s_xb_y(E'_T+2\tilde H'_T)/M]$  (middle) and the quadrupole contribution $s_xS_yb_{x}b_{y}\tilde H''_T/M^2$ (right) for $\hat x$-polarized quarks in a nucleon transversely polarized in the $\hat y$ direction. The upper (lower) row gives the results for up (down) quarks.}
\label{fig:fig5}
\end{figure}

\begin{figure}[h]
\begin{center}
\includegraphics[width=16 cm]{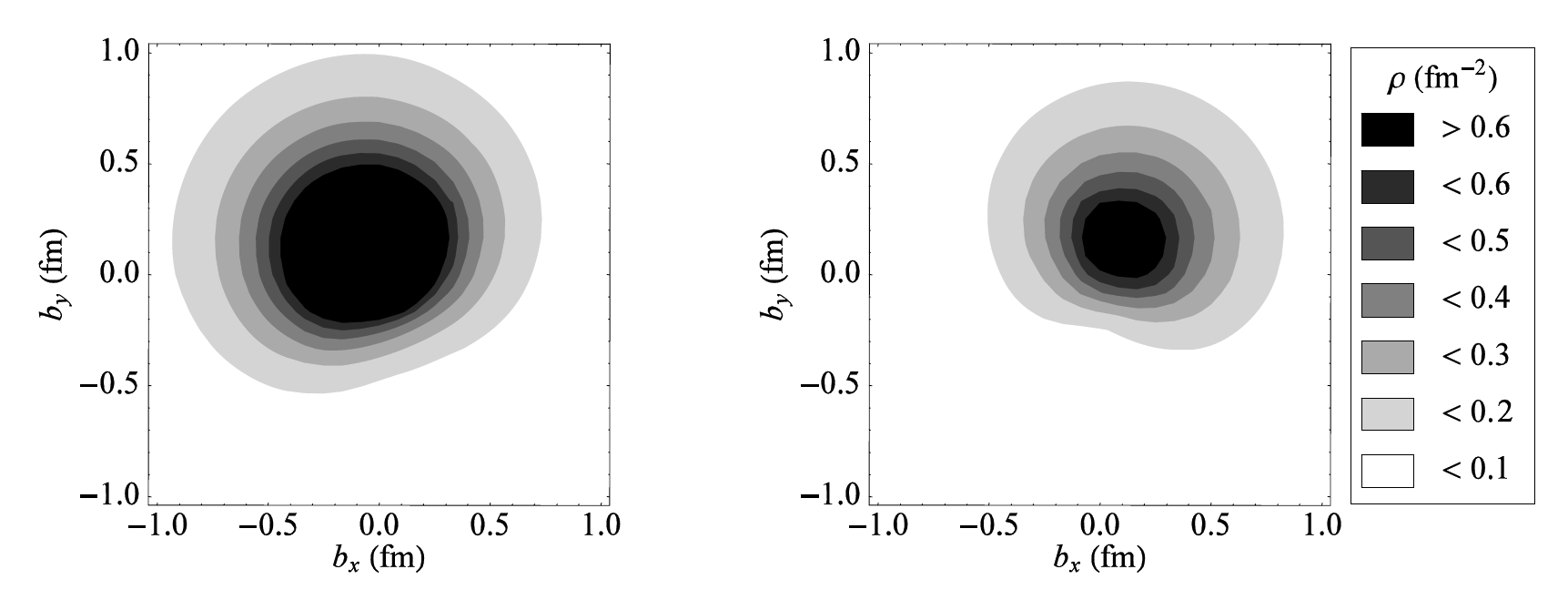}
\end{center}
\caption{\small The total spin distribution, as a sum of monopole, dipole and quadrupole terms, for $\hat x$-polarized quarks in a proton transversely polarized in the $\hat y$ direction; left (right) panel for up (down) quarks.}
\label{fig:fig6}
\end{figure}

\begin{figure}[h]
\begin{center}
\includegraphics[width=16 cm]{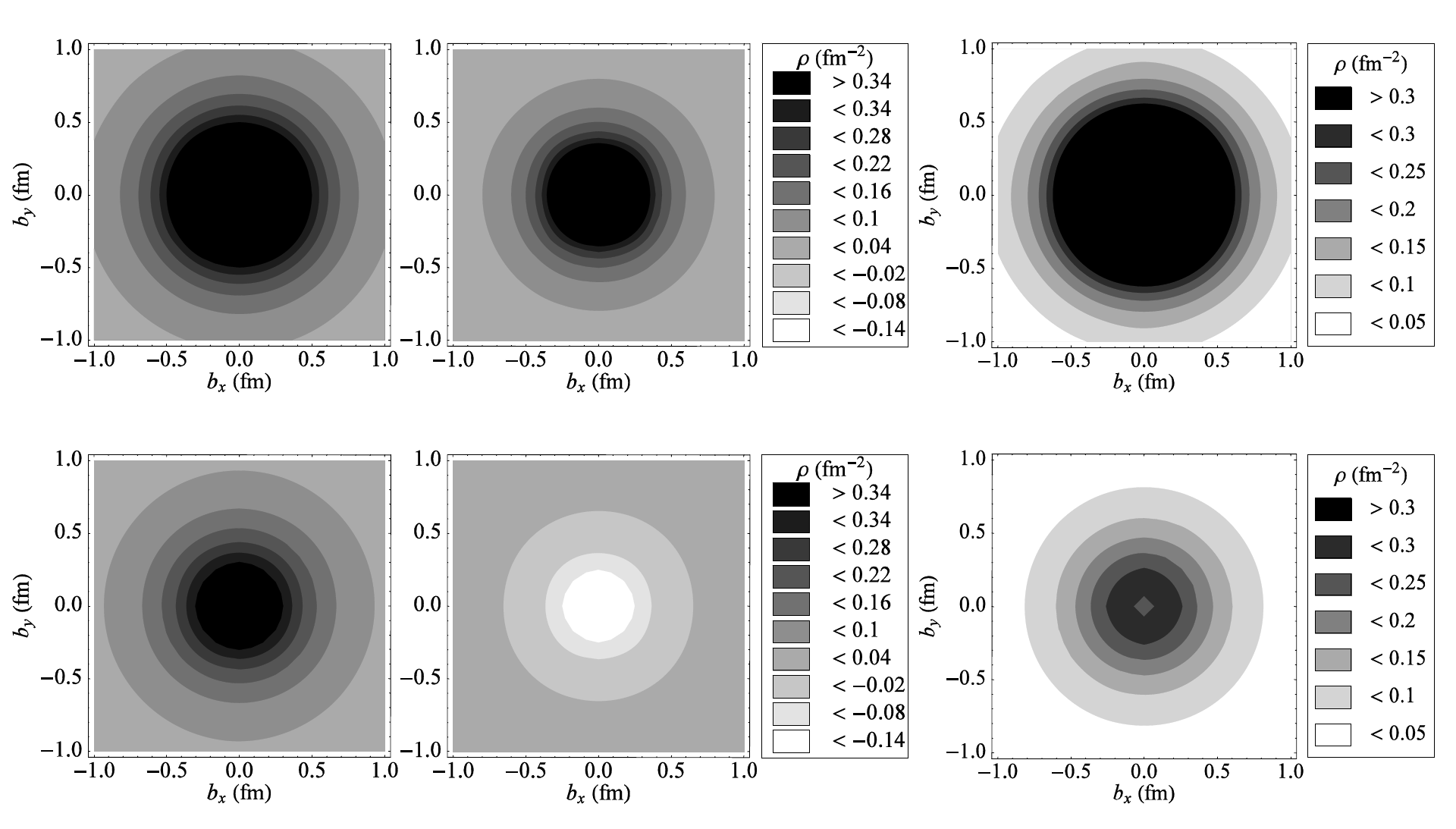}
\end{center}
\caption{\small The monopole contribution $\frac{1}{2}H$ (left) and $\frac{1}{2}\tilde H$ (right), and their sum corresponding to the spin distribution for quark polarized in the longitudinal direction, parallel to the  
proton helicity. The upper (lower) row gives the results for up (down) quarks.}
\label{fig:fig7}
\end{figure}

All the different structures appearing in Eqs.~(\ref{eq:long}) and (\ref{eq:transv}) are discussed in the following by considering the results shown in Figs.~\ref{fig:fig1} to \ref{fig:fig7}. 

In Figs.~\ref{fig:fig1} and \ref{fig:fig2} the distorting effect of the dipole 
terms due to the transverse spin distributions on the monopole terms 
corresponding to  spin densities for unpolarized quarks in an unpolarized 
target is shown. 
If one multiplies the up and down monopole terms by the quark charge $e_q$ 
and sums over quark flavors, one obtains the nucleon parton charge density in 
transverse space. 
In particular, we find that the central charge distribution for neutron 
target is negative, in agreement with the observations of Ref.~\cite{Miller07}.

For transversely polarized quarks in an unpolarized proton the dipole contribution introduces a large distortion perpendicular to both the quark spin and the momentum of the proton (Fig.~\ref{fig:fig1}). Evidently, quarks in this situation also have a transverse component of orbital angular momentum. This effect has been related~\cite{Burkardt05b,Burkardt07} to a nonvanishing Boer-Mulders function~\cite{BM98} $h_1^\perp$ which describes the correlation between intrinsic transverse momentum and transverse spin of quarks. Such a distortion reflects the large value of  the anomalous tensor magnetic moment $\kappa_T$ for both flavours. Here, $\kappa^u_T=3.98$ and $\kappa^d_T=2.60$, to be compared with the values $\kappa^u_T\approx 3.0$ and $\kappa^d_T\approx 1.9$ of Ref.~\cite{QCDSF06a} due to a positive combination $E_T+2\tilde H_T$. Since $\kappa_T\sim - h_1^\perp$, the present results confirm the conjecture that $h_1^\perp$ is large and negative both for up and down quarks~\cite{Burkardt05b,Burkardt07}.

As also noticed in Refs.~\cite{Burkardt00a,QCDSF06a} the large anomalous magnetic moments $\kappa^{u,d}$ are responsible for the dipole distortion produced in the case of unpolarized quarks in transversely polarized nucleons (Fig.~\ref{fig:fig2}). With the present model, $\kappa^u=1.86$ and $\kappa^d=-1.57$, to be compared with the values $\kappa^u=1.673$ and $\kappa^d=-2.033$ derived from data. This effect can serve as a dynamical explanation of a nonvanishing Sivers function~\cite{Siversa} $f_{1T}^\perp$ which measures the correlation between the intrinsic quark transverse momentum and the transverse nucleon spin. The present results, with the opposite shift of up and down quark spin distributions imply an opposite sign of $f_{1T}^\perp$ for up and down quarks~\cite{Burkardt02,Burkardt04a} as confirmed by the recent observation of the HERMES collaboration~\cite{Hermes05a}.

The results in Figs.~\ref{fig:fig1} and \ref{fig:fig2} are in qualitative agreement with those obtained in lattice calculations~\cite{QCDSF06a}, where
strongly distorted spin densities for transversely polarized quarks in an unpolarized nucleon have been found.  
One observes that the sideways distortion for down quarks is about twice as strong as for up quarks, even if the anomalous magnetic moment $\kappa^q$ and the anomalous tensor magnetic moment $\kappa_T^q$ have about the same magnitude. 
This is because the monopole distribution for up quarks is twice as large as 
for down quarks, and therefore adding the dipole contribution results in a 
larger distortion for down quarks than for up quarks.

In Fig.~\ref{fig:fig3} quarks and proton all transversely polarized along 
$\hat x$ are considered. Quite remarkably, opposite 
signs of the up and down quadrupole and spin dependent monopole  
terms are found as a 
consequence of the opposite sign for up and down quarks 
 of the $x$ dependence of $H_T$ and  $\tilde H_T$ predicted by the model~\cite{PPB05}.
The quadrupole distribution for up quark is more spread than for down quark,
 but it is much smaller, resulting in an average 
distortion equal to -0.04 to be compared with the value 0.07 for down quark.

The total spin distribution for quarks and proton transversely polarized along $\hat x$ is shown in Fig.~\ref{fig:fig4} as the result of summing for each flavour the two monopole contributions in the left panels of Figs.~\ref{fig:fig1} and \ref{fig:fig3}, the two dipole contributions on the middle panels of Figs.~\ref{fig:fig1} and \ref{fig:fig2} and the quadrupole contribution of the right panel in Fig.~\ref{fig:fig3}. For up quarks this distribution is almost axially 
symmetrical around a position slightly shifted in the $\hat y$ direction (Fig.~\ref{fig:fig4}). This is a consequence of the dominating role of the monopole terms, the compensating action of the two dipole terms and the small quadrupole contribution. In contrast the down quark spin distribution has a much lower size and 
shows a strong and symmetric deformation about the $\hat y$ axis stretching 
along the same $\hat x$ direction of the quark and proton polarization. 

The resulting transverse shift of both up and down distributions  in Fig.~\ref{fig:fig4} is suggesting the presence of an effective transverse quark orbital angular momentum introduced in the LCWFs by the Melosh rotation required to transform the canonical spin to the light-front spin. Due to the shift in the positive $\hat y$ direction transverse quark spin and transverse quark orbital angular momentum seem to be aligned along the same $\hat x$ direction of the proton polarization.

In Figs.~\ref{fig:fig5} and \ref{fig:fig6} results are given for $\hat x$-polarized quarks in a proton polarized along $\hat y$. The distortion due to the dipole contribution $\oneh S_yb_xE'$ in Fig.~\ref{fig:fig5} is rotated with respect to the case shown in Fig.~\ref{fig:fig1}, but the origin of opposite shift for up and down quarks is always the opposite sign of the anomalous magnetic moments $\kappa^{u,d}$. The total dipole distortion in Fig.~\ref{fig:fig5} is obtained by considering also the second dipole term $-\oneh s_xb_y(E'_T+2\tilde H'_T)/M$ displayed in Fig.~\ref{fig:fig1}. The result is quite sizable, while the quadrupole term rather small. Therefore, the total resulting distortion of the spin density (Fig.~\ref{fig:fig6}) is  due to the dipole terms with a small contribution from the quadrupole terms. Correspondingly, the quark orbital angular momentum has positive $\hat x$ and $\hat y$ components for up quarks, and positive $\hat x$ and negative $\hat y$ components for down quarks. Here as well as in 
Fig.~\ref{fig:fig4} the quark orbital angular momentum is entirely generated by the
 Melosh rotations.

Finally, the case of quark polarization parallel to the proton helicity is considered in Fig.~\ref{fig:fig7}. Here only monopole terms occur (see Eq.~(\ref{eq:long})) and their role was first discussed in Ref.~\cite{Burkardt03}. The opposite sign of $\oneh \tilde H$ for up and down quarks is responsible for quite a different radial distribution of the axially symmetric spin density. Since in the forward limit the GPD $\tilde H$ reduces to the helicity distribution $g_1(x)$, $\tilde H(x,\xi=0,t=0)=g_1(x)$, this difference ultimately reflects the opposite behaviour of the helicity distributions and the opposite sign of the axial-vector coupling constants $g_A^{u,d}$ of up and down quarks (see also Ref.~\cite{BPT04}).

In the present analysis only distributions for $x>0$ have been considered. In principle, also $x<0$ and antiquarks contributions should be considered. However, at the hadronic scale of the model contributions from antiquarks are expected to be small in general. For the transverse distributions one also knows that gluons do not mix under evolution. Therefore their qualitative behaviour presented here in agreement with results from lattice calculations~\cite{QCDSF06a} can be considered sufficiently indicative. In contrast, sea quarks and gluons can affect longitudinal distributions under evolution. In this case, the present results could be improved by considering the meson cloud surrounding the bare nucleon at the hadronic scale and including its contribution in the evolution~\cite{PB06}.




\end{document}